# Characterization of exhaled e-cigarette aerosols in a vape shop using a field-portable holographic on-chip microscope

*Ege Çetintaş[1,2,3]*    email: egecetintas1@g.ucla.edu

*Yi Luo[1,2,3]*    email: yluo2016@ucla.edu

*Charlene Nguyen[4]*    email: charchar626@gmail.com

*Yuening Guo[4]*    email: ynguo94@g.ucla.edu

*Liqiao Li[4]*    email: liqiao93@g.ucla.edu

*Yifang Zhu[4]*    email: yifang@ucla.edu

*Aydogan Ozcan[1,2,3,5,\*]*    email: ozcan@ucla.edu

[1]Electrical and Computer Engineering Department, University of California, Los Angeles, California 90095, USA

[2]Bioengineering Department, University of California, Los Angeles, California 90095, USA

[3]California Nano Systems Institute (CNSI), University of California, Los Angeles, California 90095, USA

[4]Department of Environmental Health Sciences, University of California, Los Angeles, California 90095, USA

[5]David Geffen School of Medicine, University of California, Los Angeles, California 90095, USA

[*]Correspondence: Prof. Aydogan Ozcan

E-mail: ozcan@ucla.edu

Address: 420 Westwood Plaza, Engr. IV 68-119, UCLA, Los Angeles, CA 90095, USA

# Abstract

The past decade marked a drastic increase in the usage of electronic cigarettes (e-cigs). The adverse health impact of secondhand exposure due to exhaled e-cig particles has raised significant concerns, demanding further research on the characteristics of these particles. In this work, we report direct volatility measurements on exhaled e-cig aerosols using a field-portable device (termed c-Air) enabled by deep learning and lens-free holographic microscopy; for this analysis, we performed a series of field experiments in a vape shop where customers used/vaped their e-cig products. During four days of experiments, we periodically sampled the indoor air with intervals of ~15 minutes and collected the exhaled particles with c-Air. Time-lapse inline holograms of the collected particles were recorded by c-Air and reconstructed using a convolutional neural network yielding phase-recovered microscopic images of the particles. Volumetric decay of individual particles due to evaporation was used as an indicator of the volatility of each aerosol. Volatility dynamics quantified through c-Air experiments showed that indoor vaping increased the volatility of particles as well as the percentage of volatile and semi-volatile particles in air. The reported methodology and findings can guide further studies on volatility characterization of e-cig emission and regulations on indoor vaping.

**Introduction:**

Electronic cigarettes (e-cigs) experienced wide spread use in the past few years with never-smoking adolescents and young adults being the prominent consumer base[1,2]. These small handheld devices generally vaporize e-cigarette liquids (e-liquids) that contain nicotine and flavorings dissolved in Propylene Glycol (PG) and Vegetable Glycerin (VG), with various volumetric ratios of these two chemicals[3]. The usage of an e-cig (or 'vaping') produces a cloud of rapidly evaporating particles. The potential adverse health effects due to inhalation of these particles triggered numerous research studies[4]. Laboratory-based studies revealed that the particle concentration, size and mass distribution and chemical composition of the emitted aerosols are linked to the e-liquid type[5–9], power ratings of the vaporizer[10–13] and environmental factors like room temperature and humidity[14–17]. Recent studies also measured the dynamic changes of e-cig emissions by measuring particle volatility using a portable computational imaging device, termed c-Air[18]. The c-Air device[18,19] collects e-cig emitted particles using an impactor-based sampler and a miniaturized pump. In-line holograms[20–22] of collected particles are recorded at a frame rate of 2 fps (frames/sec) and these time-lapse holograms are further processed using a trained neural network[23,24] that simultaneously performs auto-focusing and phase recovery. The recovered microscopic images are used to estimate each particle's volume as a function of time thus enabling us to quantify the volatility of particles. These previous c-Air based studies were performed in laboratory settings (without any human vaping) and revealed that the volatility of e-cig particles changed as a function of the chemical composition of the e-liquids such as the volumetric ratio of PG and VG, as well as the amount of nicotine found in the e-liquid[18].

In addition to these laboratory-based experiments, several additional studies focused on the exhaled e-cig aerosols inside vape shops[25,26]. Vape shops are retail establishments, where e-cig products are sold and consumed. Different from in-lab studies that focus on aerosols directly emitted from an e-cig device, e-cig aerosols found in vape shops are exhaled particles that went through human lungs (also known as secondhand smoke). Compared to in-lab studies, the dynamics of e-cig aerosols in vape shops are much more complex due to human vaping, the usage of various e-cig products and the presence of non-homogenous air in the room. Additionally, less-controlled environmental variables such as the room temperature and humidity add other factors of complexity to the dynamics of these aerosols.

In this work, we report direct volatility measurements on exhaled e-cig aerosols in a vape shop using the c-Air device (Fig. 1(a)). In a randomly selected vape shop in Los Angeles, USA, we performed periodic sampling of the indoor air quality every ~15 min, and analyzed the exhaled particles within the vape show with c-Air device, which recorded time-lapse inline holograms of the detected aerosols. Using the phase information channel of the reconstructed holograms, the volumetric decay of each captured particle was measured as a function of time, inferring the volatility information of e-cig aerosols within the vape shop. Through these volatility dynamics measured with our c-Air device in the vape shop, we showed that indoor vaping resulted in an increase in the particle volatility as well as an increase in the percentage of volatile and semi-volatile particles in air. Our results and analyses highlight the complex temporal dynamics of e-cig related particle emission within indoor spaces, and the presented method can be used to guide regulations on indoor vaping and secondhand smoking.

**Results and Discussion**:

The field experiments were conducted in a vape shop that was randomly selected (refer to the Methods section for details). This vape shop occupies a single room with an employee-only area on one side of the room. Approximately two meters away from where customers and employees vaped (Fig. 1(b)), the c-Air device was placed on the display countertop to capture e-cig generated aerosols. In addition to the c-Air device used within the vape shop, an Aerodynamic Particle Sizer (APS) was configured to take samples every two minutes to provide an independent measurement of the particle size distribution within the vape shop.

During these field experiments, c-Air was controlled by our team to collect air samples, with additional, on-demand measurements done upon the observation of vaping activities. The time between two consecutive c-Air measurements $\Delta T_k$ was kept between 5 to 30 minutes (Fig. 1(c)) with an average of ~16 minutes. Occasionally, multiple customer vaping events (e.g., continuous e-cig puffing in short periods of time) occurred between two consecutive c-Air measurements and these time instances were marked for further analysis. During each measurement step, c-Air sampled the indoor air for 60 seconds (sampling ~13L of air) and simultaneously recorded time-lapse inline holograms of the collected particles at 2 fps. After the initial 60-second time window, the vacuum pump stopped and the hologram recording continued for an additional 30 seconds, totaling 90 seconds of imaging time at 2 fps. A deep neural network[18,23] was designed to reconstruct the captured time-lapse holograms (Fig. 2(a)), providing 180 complex-valued reconstructed images, each with an independent amplitude and phase channel. At each time frame, the individual particles were detected via thresholding of both the amplitude and phase channels with an adaptive threshold of five standard deviations above the mean value of each image. The union of these binary spatial masks from both the amplitude and the phase channels provides an approximate boundary for each detected particle's region-of-interest (ROI) at each time point. Having detected a collection of ROIs for all the captured aerosols, the particle volume at any time point *t* (see e.g., Supplementary Videos 1-3) was estimated by integrating the phase values ($\phi$) within each one of these ROIs, i.e.,

$$V(t) = \frac{\lambda}{2\pi\Delta n} \cdot \sum_{i,j} \phi(I(i,j;t)) \quad (1)$$

where we assumed a refractive index difference of $\Delta n = 0.4$ with respect to air/vacuum; this is a reasonable assumption as it reflects the typical refractive index value for PG and VG[27,28]. In this equation, $\lambda$ is the illumination wavelength (850 nm) and *I* is the reconstructed complex image of each particle within the ROI. The summation in Eq. (1) is carried over a set of lateral pixels {i,j} defined by an additional spatial mask generated specifically for the particles of interest. These binary masks were created by taking the pixels that are at least one standard deviation above the mean value of the ROI.

Based on Eq. (1), the volume decay of each aerosol was approximated using an exponential fit:

$$V(t) = \alpha e^{-\kappa t} + b \quad (2)$$

In Eq. (2), the exponential term was optimized to fit the volume decay of the particle, whereas the second term was optimized to match any DC term, representing some of the particles that do not fully evaporate (see e.g., Supplementary Videos 2 and 3). The lifetime of the particle, $\tau$, is further defined as

$$\tau = \frac{1}{\kappa} \quad (3)$$

To be able to quantify particles with different sizes, the initial volume of a particle, $V(t_0)$, right after it lands on the collection substrate of c-Air is divided by its corresponding lifetime $\tau$:

$$\Delta V_\tau = \frac{V(t_0)}{\tau} \tag{4}$$

giving us the *volume decay rate* $\Delta V_\tau$ (µm³/sec) which functions as a figure of merit for the volatility of an evaporating particle (Fig. 2(c)).

During the four days of c-Air experiments in the vape shop, we conducted a total of 115 different measurements, whereas the APS sampled the air at a much faster rate of one sample per two minutes. The particle emission due to human vaping is depicted by the sharp rises in particle concentration measured by APS (see the red bars in Fig. 3). To reveal the dynamic changes of all the collected particles (usually on the order of hundreds to thousands of particles per c-Air measurement), the average volume decay rate of each measurement was calculated and plotted in Fig. 3, blue dotted lines. The volume decay rate (µm³/sec) quantifies a particle's evaporation speed, and therefore higher volume decay rates indicate faster evaporation and higher volatility. The correlation that is observed in Fig. 3 between the dynamics of the mean volume decay rate and the APS measurements reports a link between the e-cig emissions of vapers and aerosol volatility (Fig. 3). Also note that the local minimums and maximums of the c-Air measured volume decay rates reported in Fig. 3 for different measurement days match the corresponding local extreme points in our APS measurements.

Given the complex chemical composition of the captured e-cig aerosols (from various e-liquids), we defined three categories of aerosols according to the volumetric evolution of the imaged particles: volatile, semi-volatile and non-volatile particles (see Fig. 4 and Supplementary Videos 1-4). A particle is defined to be volatile if it fully evaporates (i.e., its volume gets smaller than an empirical threshold of 0.1 μm$^3$ before the end of each measurement) and its volume exhibits a smooth exponential decay (see e.g., Supplementary Video 1). Since the c-Air device records an additional set of 60 frames after the air sampling is complete, all the volatile particles that were captured by c-Air fully evaporated within our observation time window. The mean and standard deviation of the volume decay rates of all the volatile particles captured in 115 measurements were used to further differentiate the remaining particles, forming an empirical definition for semi-volatile particles, i.e., they exhibit an initial volume decay, followed by a second phase where they remain larger than 0.1 μm$^3$ in volume. Semi-volatile particles can be coagulated particles or particles with solid cores[10,29] (see e.g., Supplementary Videos 2 and 3). The remaining particles that did not exhibit detectable evaporation were defined as non-volatile particles. These particles exhibited a very small volume decay rate which was empirically found to be less than 0.05 μm$^3$/sec. For these non-volatile particles, the subtle changes in their measured volumes as a function of time might be related to the deformation of the sampling pad after the particle's impaction (see Supplementary Video 4). For each c-Air measurement point, the percentage of volatile and semi-volatile particles combined is color-coded and shown in Figure 3. The change in the ratio of the volatile particles to other detected particles was observed to be similar to the time dynamics of the volume decay rate that we measured using c-Air (see the colorbar in Fig. 3 for each measurement day). This indicates that the percentage of the volatile particles within the detected aerosols can be used an indirect measure of aerosol volatility and volume decay rate.

To further expand our analysis, next we focus on 3 measurements, i.e., a, b and c marked in Fig.

3, that were conducted while no human vaping was observed in the vape shop. Given the dynamic indoor environmental conditions throughout a day, in terms of e.g., temperature, humidity and opening/closing of the outside door, these vape-free measurements within the vape shop can be considered as a background state for comparison purposes. An increase in aerosol volatility and volatile particle ratio can be clearly seen in the following 3 subsequent measurements points a', b' and c' (see Fig. 3) during which the customers in the shop vaped, significantly increasing the e-cig generated aerosols. Furthermore, dense and rapid vaping events within the vape shop also increased the aerosol volatility that we measured. For example, a continuous burst of vaping was observed close to our c-Air device on March $9^{th}$, at around 14:30 PM (Fig. 3). As a result of this continuous vaping within the shop, a sharp rise in $PM_{10}$ emission, particle volatility as well as volatile particle ratio is clearly observed in Fig. 3. In this case, c-Air measurements revealed that >95% of the particles fell in the category of volatile particles and the measured volume decay rates were also notably higher compared to other measurement points within the same day, which shows how dense, successive vaping events can significantly increase the aerosol volatility within a room.

In conclusion, we conducted field experiments in a vape shop to characterize volatility of e-cig emission using a field-portable, high-throughput device which can sample aerosols at a rate of 13 L/min and continuously image the collected particles on an impaction-based sampler. These field experiments were carried out in a vape shop where the customers and employees vaped their own e-cigs, resulting in the vaporization of e-liquids with different chemical constituents. The c-Air device enabled us to image these microscale particles generated by e-cigs during their evaporation lifetimes, allowing us to quantify their numbers, size and volatility. These field experiments revealed that vaping in the vape shop resulted in an increase in the particle volatility as well as an increase in the percentage of volatile and semi-volatile aerosols.

## Methods

**Vape shop selection**

The vape shop was randomly selected from 67 candidate vape shops in Los Angeles County, USA. The candidate list was generated by a Yelp search of 'vape shop', while only keeping the stores that provide *solely* e-cigs. This selected vape shop, located in Santa Clarita, California (USA), has a total store volume of 205 m$^3$ and is in a multi-unit plaza with a central ventilation that was not in use during our experiments.

**Portable holographic microscope for aerosol collection and quantification**

The c-Air device captures the aerosols in the vape shop and records time-lapse inline holograms of the collected aerosols. A miniaturized vacuum pump inside the c-Air device creates an air stream at a flow rate of 13 L/min towards a disposable impactor (Air-O-Cell Sampling Cassette, Zefon International, Inc.). A portion of the aerosols in the air stream are collected on the sticky transparent coverslip due to inertial impaction[19]. The collected particles are illuminated using a vertical-cavity surface-emitting laser (VCSEL) diode (OPV300, TT Electronics, $\lambda_{peak} = 850\ nm$) to create in-line holograms of the deposited aerosols on the transparent substrate. The holograms are digitally recorded at 2 fps using a complementary metal–oxide–semiconductor (CMOS) image sensor chip (Sony IMX219PQ, pixel pitch 1.12 μm). A Raspberry Pi Zero W single-board computer within c-Air is used for interfacing with the CMOS image sensor, the illumination source and the vacuum pump.

**PM$_{10}$ mass concentration estimation**

An Aerodynamic Particle Sizer (APS 3321, TSI Inc.) was also used in our field tests to provide real-time measurements of particulate matter mass concentration. The APS device provides a measurement of the particle size distribution, covering the particles ranging from 0.5 μm to 19.8 μm. $PM_{10}$ (i.e., particulate matter with an aerodynamic diameter of ≤ 10 μm) mass concentration was estimated using APS data in the size range of 0.5-10 μm by assuming spherical particles with a density of 1.1 g/cm$^3$.[30,31]

# List of Figures

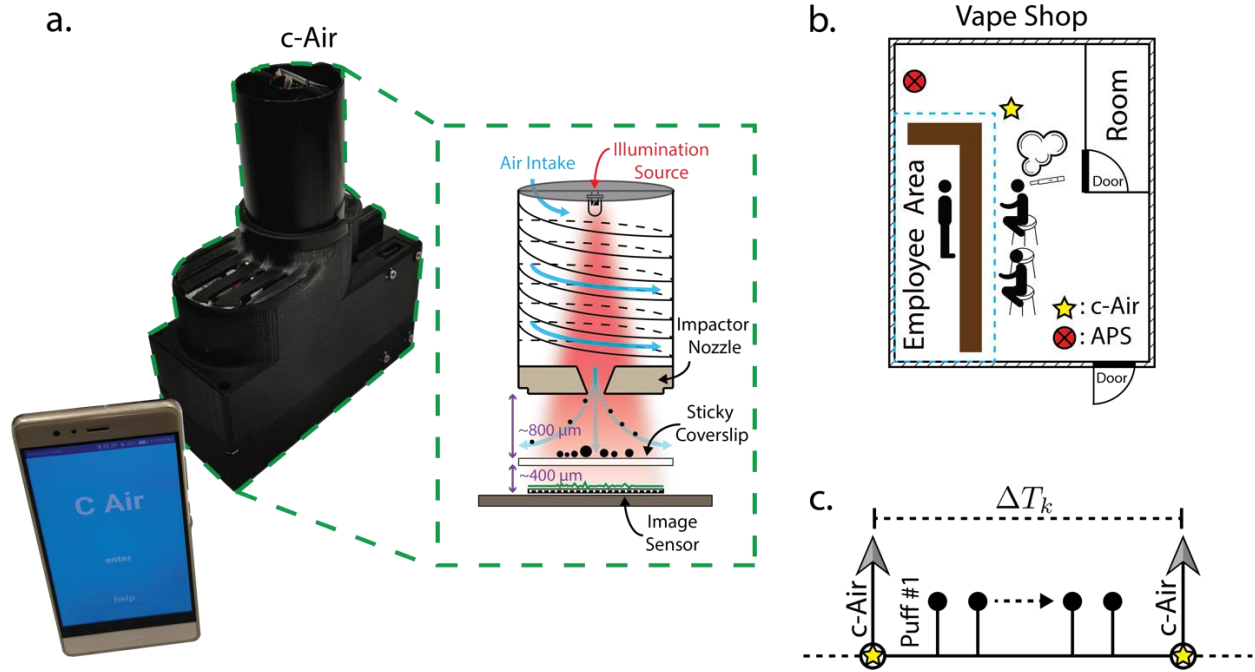

**Figure 1. The c-Air device and the field experiments in a vape shop for aerosol volatility measurements. a.** A photograph of the c-Air device and the mobile phone application to interface with the device. The schematic of the c-Air device. **b.** Floor plan of the vape shop, marking the measurement devices and e-cig users. **c.** A schematic timing diagram of vaping events and c-Air measurements.

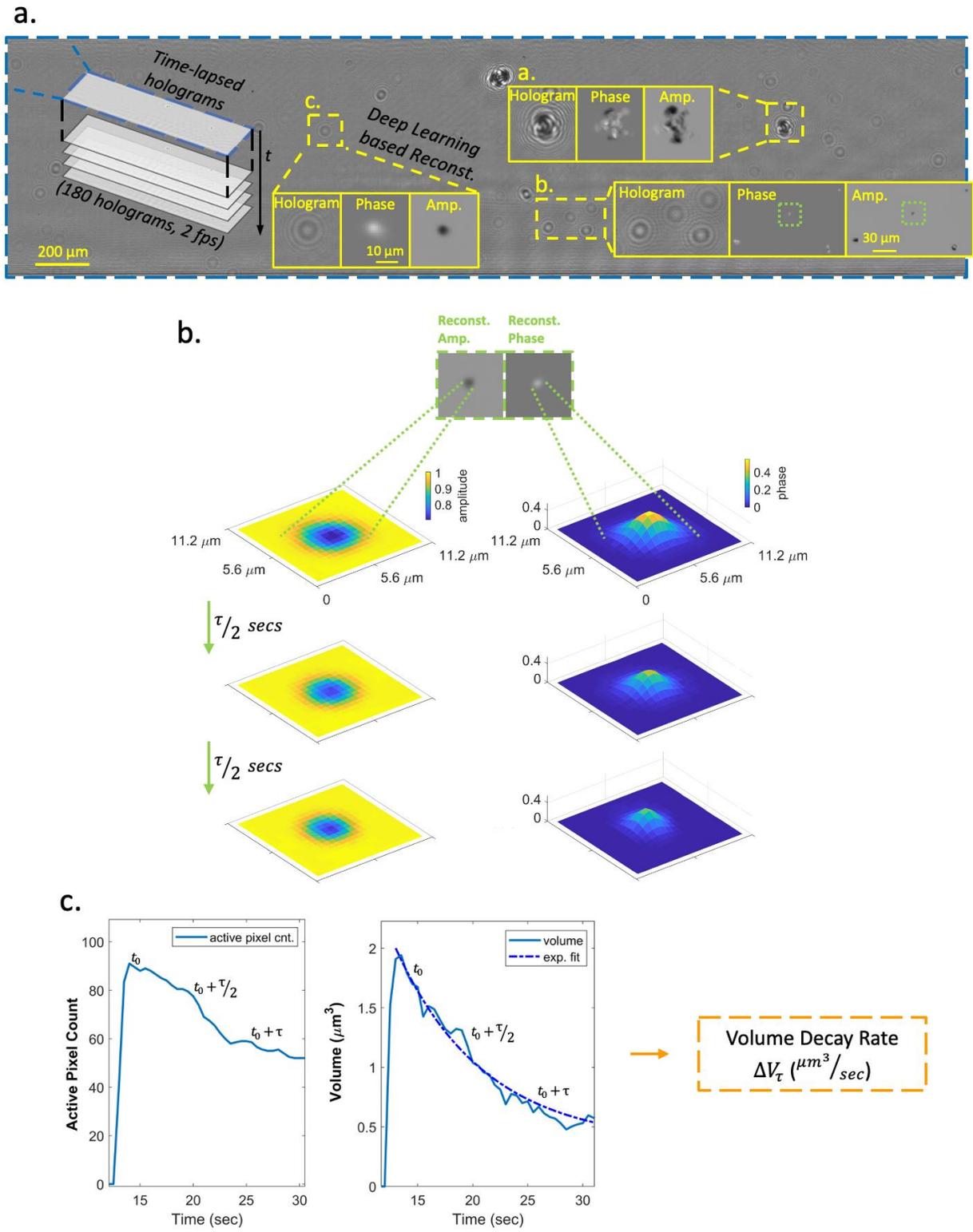

**Figure 2. Calculation of the volatility of each detected aerosol: from raw holograms to volume decay rate. a.** A full field-of-view raw inline hologram and some example

reconstructions showing the phase and amplitude channels separately. **b.** An example aerosol image and the evolution of its holographic phase and amplitude channels at time points 0, $\tau/2$ and $\tau$, where $\tau$ is the exponential time constant (see Eq. (3)). **c.** The plot of the active pixels in the amplitude channel as a function of time (left) and the exponential decay of the particle volume (right).

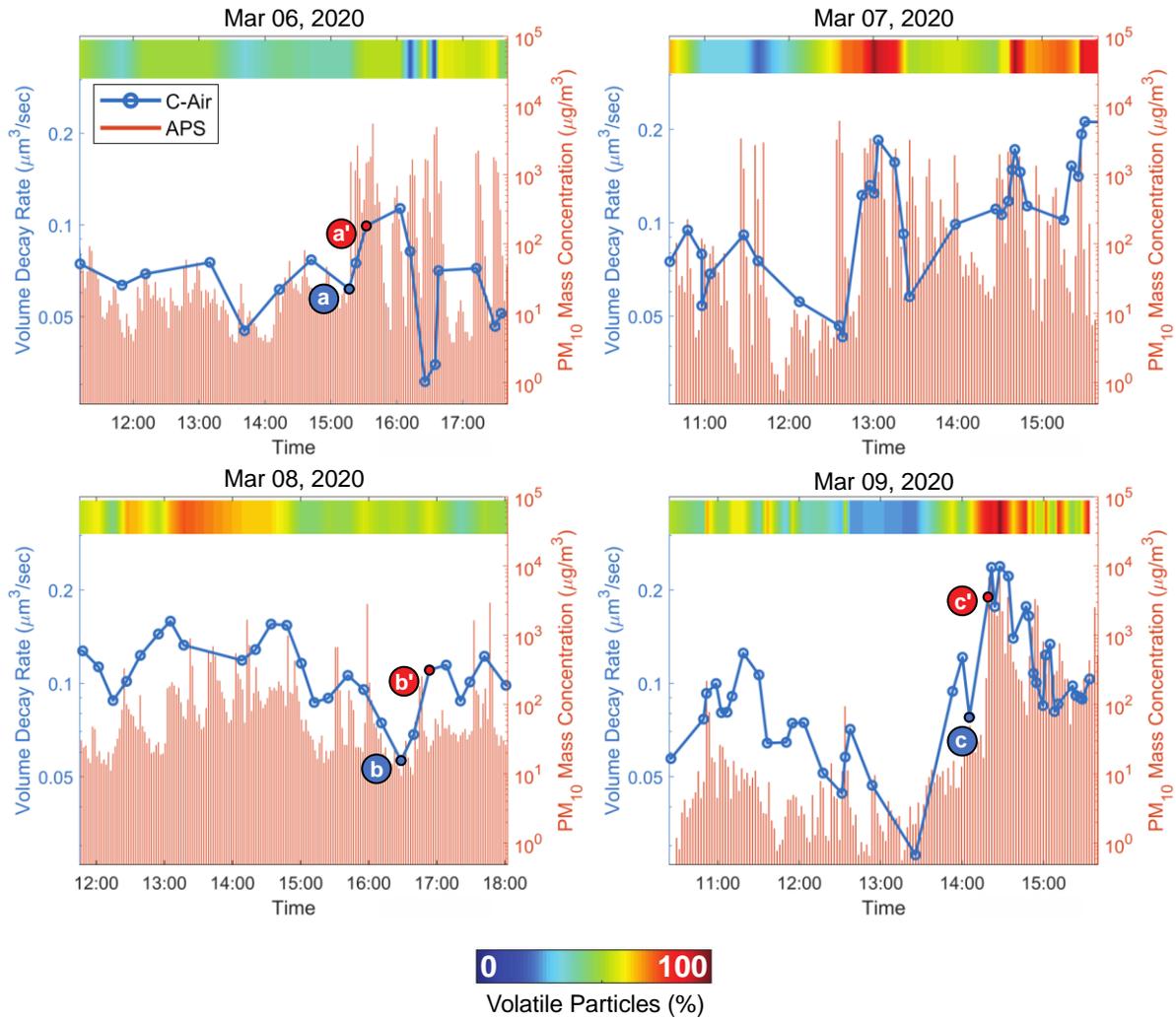

**Figure 3. The change of the mean volume decay rate and the percentage of volatile aerosols throughout different days of experiments.** *a*, *b* and *c* mark the data points where there was no observation of a vaping event prior to c-Air sampling, whereas, *a′*, *b′* and *c′* mark the measurement points where vaping was observed in the vape shop.

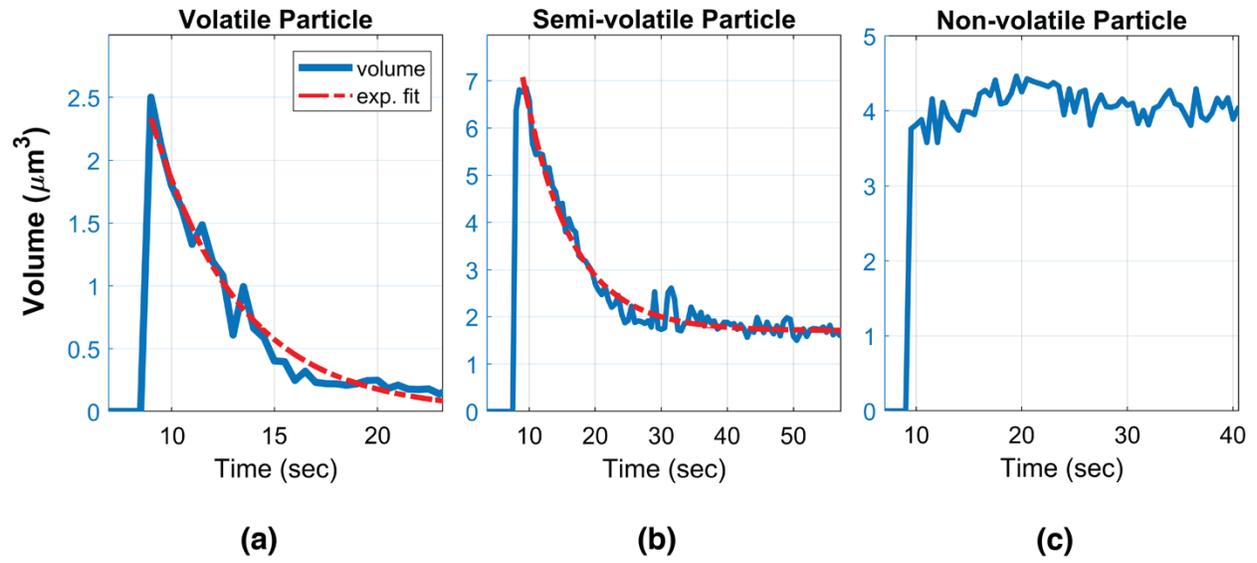

**Figure 4.** Characteristic volume change of a (a) volatile, (b) semi-volatile and (c) non-volatile particle, sampled within the vape shop.